\documentclass[twocolumn,10pt,english,aps,prl,longbibliography]{revtex4-2}
\usepackage[T1]{fontenc}
\usepackage[utf8]{inputenc}
\usepackage{color}
\usepackage{babel}
\usepackage{verbatim}
\usepackage{titlesec}
\usepackage[dvipsnames]{xcolor}
\usepackage{amsbsy}
\usepackage{amsmath}
\usepackage{amstext}
\usepackage{graphicx}
\usepackage{afterpage}
\usepackage{placeins}
\usepackage{enumerate}
\usepackage{multirow}
\usepackage{amssymb}
\usepackage{units}
\usepackage[unicode=true,pdfusetitle,
bookmarks=true,bookmarksnumbered=false,bookmarksopen=false,
 breaklinks=false,pdfborder={0 0 0},pdfborderstyle={},backref=false,colorlinks=true]{hyperref}
\hypersetup{urlcolor=blue,citecolor=blue}

\newcommand{\noop}[1]{}



\begin{document}

\title{Electronic band structure reconstruction in Ni intercalated  ZrTe$_{2}$}

\author{Pedro H. A. Moya\texorpdfstring{$^{1,2}$}{1,2†},
Marli R. Cantarino\texorpdfstring{$^{3}$}{3},
Lauro B. Braz\texorpdfstring{$^{1}$}{1},
Lucas E. Correa\texorpdfstring{$^{4}$},
Leandro R. de Faria\texorpdfstring{$^{4}$},
Rodrigo M. C. Huamani\texorpdfstring{$^{1}$}{1},
Wendell S. Silva\texorpdfstring{$^{4}$}{4},
Claude Monney\texorpdfstring{$^{6}$}{6},
Antonio J. S. Machado\texorpdfstring{$^{5}$}{5},
and Fernando A. Garcia\texorpdfstring{$^{1}$}{1†}}

\affiliation{$^{1}$Instituto de Física, Universidade de São Paulo, 05508-090 São Paulo, SP, Brazil}
\affiliation{$^{2}$Department of Materials Science and Applied Mathematics, Malmö University, 211 19 Malmö, Sweden}
\affiliation{$^{3}$European Synchrotron Radiation Facility, BP 220, F-38043 Grenoble Cedex, France}
\affiliation{$^{4}$Brazilian Synchrotron Light Laboratory (LNLS), Brazilian Center for Research in Energy and Materials (CNPEM), Campinas-SP, 13083-970, Brazil}
\affiliation{$^{5}$Universidade de São Paulo, Escola de Engenharia de Lorena, DEMAR, Lorena, Brazil}
\affiliation{$^{6}$Département de Physique, Université de Fribourg, CH-1700 Fribourg, Switzerland}

\begin{abstract}

The filling of the large van der Waals gap in Transition Metal Dichalcogenides (TMDs) often leads to lattice and electronic instabilities, which prelude the onset of a rich phenomenology. Motivated by previous results from macroscopic measurements, we investigate the electronic structure of the TMDs ZrTe$_2$ and Ni intercalated ZrTe$_2$ (Ni$_x$ZrTe$_2$, $x\approx 0.05$) employing angle-resolved photoemission spectroscopy (ARPES). We readily identify in Ni$_x$ZrTe$_2$ two flat bands, most likely associated with localized Ni-derived 3$d$-states, at about $\approx-0.7$ eV and $\approx-1.2$ eV in binding energy.  More significantly, at low $T$, we identify an electronic structure reconstruction in Ni$_x$ZrTe$_2$, which halves the electronic periodicity along the $k_{z}$ direction. This is reminiscent of a commensurate band folding with wave-vector $q=(0,0,\pi)$. A mean-field model for a charge density wave (CDW) transition driven by interband hole-hole interactions is solved with inputs from the experimentally determined electronic band structure. Comparison between theory and experiments unveils a CDW transition in Ni$_x$ZrTe$_2$, for which the gap opens above the Fermi level. Our findings invite  further investigation into the structural properties of intercalated  ZrTe$_2$. 

\end{abstract}

\maketitle

\section{\label{sec:intro}INTRODUCTION}

Layered solid-state materials are prototypical examples of quasi-two-dimensional systems, in which reduced dimensionality and lattice anisotropy play a central role in shaping the electronic ground state. These characteristics promote two-dimensional Fermi-surface topologies and enhanced susceptibility to collective electronic instabilities, providing fertile ground for the emergence of a wide range of collective phenomena. Within this broad class, transition metal dichalcogenides (TMDs) defined as $MX_{2}$, where $M$ is a transition metal and $X$ is a chalcogen, constitute one of the most extensively studied families of layered compounds. Over the past decades, TMDs have served as model systems at the intersection of fundamental condensed matter physics and materials science, owing to their tunable electronic structure, chemical versatility, and compatibility with device architectures \cite{wang_electronics_2012, han_van_2018, manzeli_2d_2017}.

Beyond their intrinsic properties, TMDs are particularly attractive because their van der Waals layered structure enables controlled chemical modification, such as intercalation, substitution, and electrostatic gating. These approaches provide powerful means to tailor charge density, lattice symmetry, magnetic interactions, and spin-orbit coupling, thereby stabilizing electronic ordered phases. Intercalation of elements, typically but not exclusively transition metals, in the van der Waals gap of TMD materials have been explored due to their potential to create and control unusual magnetic and electronic orders in TMDs.  As a result, TMD-based systems have emerged as versatile platforms to investigate charge density waves, superconductivity, magnetism, and topological states of matter \cite{zhang_kramers_2025, Wenxin_electronic_2026, shen_novel_2007, jo_electrostatically_2015, wu_discovery_2023, wu_highly_2022, pan_pressure-driven_2015, morosan_superconductivity_2006, kang_superconductivity_2015, qi_superconductivity_2016, furue_superconducting_2021, ali_large_2014}. These phenomena are not only of technological relevance but also provide insight into the fundamental mechanisms governing electronic interactions in reduced dimensions.

Within this context, dichalcogenide compounds based on tellurium ($M$Te$_2$) are of particular interest, as the large atomic mass of Te enhances spin-orbit coupling (SOC), a key ingredient for realizing nontrivial band topology. Several members of this family have therefore attracted considerable attention as potential hosts of topological electronic states. In particular, ZrTe$_2$ has recently been investigated extensively, with studies reporting unconventional electronic behavior and signatures of nontrivial band topology \cite{kar_metal-chalcogen_2020, ren_semiconductormetal_2022, muhammad_transition_2020, wang_weak_2021, wang_magnetotransport_2022, ou_zrte2crte2_2022}. These findings highlight the broader relevance of telluride-based layered materials and their chemically modified derivatives as a promising platform for exploring the interplay between dimensionality, spin-orbit coupling, and electronic order.

\begin{figure*}[t]
\includegraphics[width=\linewidth]{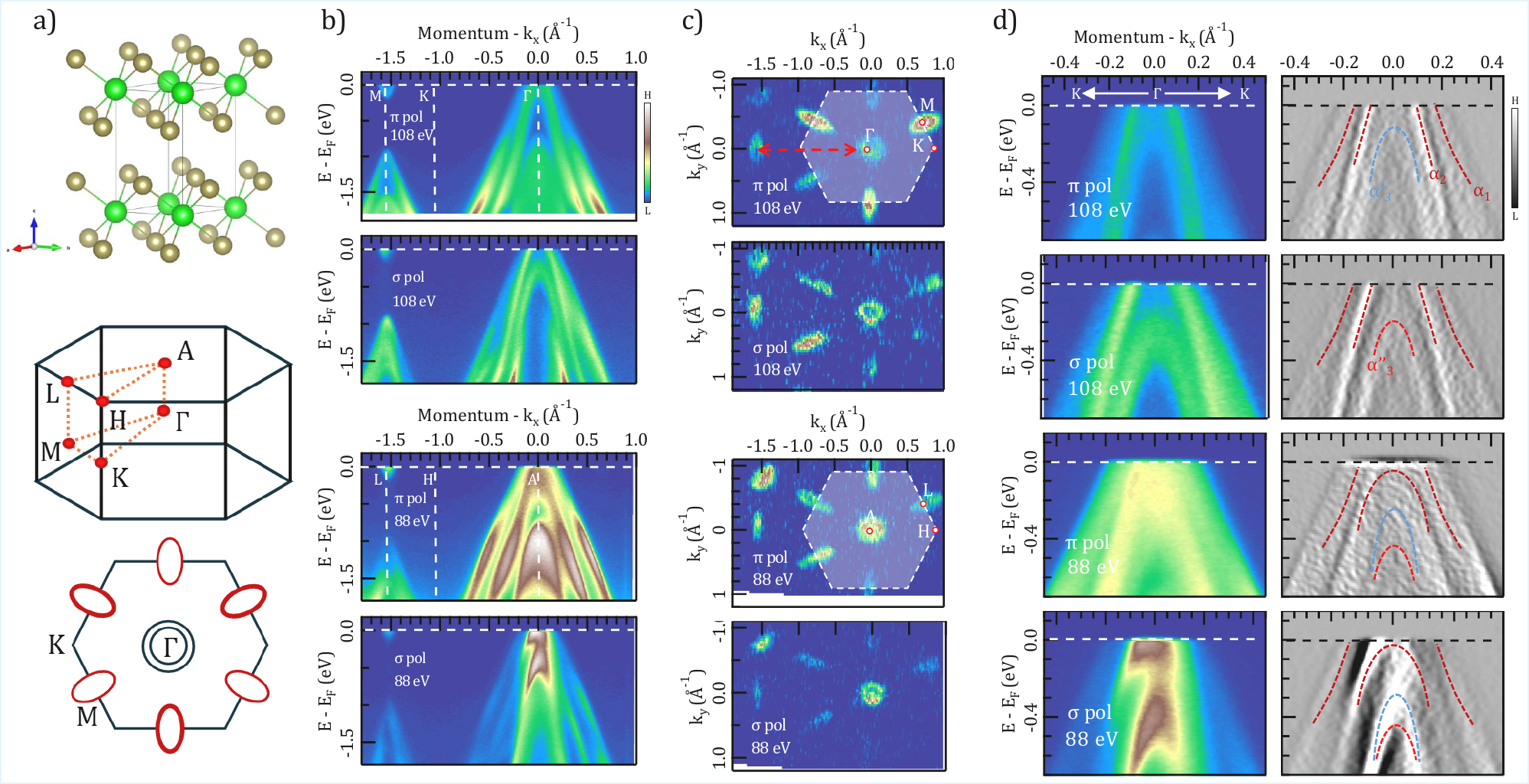}
\caption{\label{fig:1} (a) ZrTe$_2$ crystal structure (top), the respective hexagonal BZ and high symmetry points (middle) and the projection of the 2D Fermi Surface onto the $\Gamma$ plane (bottom). Zr and Te atoms are represented by green and gold spheres, respectively. (b) EDMs for both photon polarizations and BZ planes for ZrTe$_2$, measured at $T=20$ K. Lower panels display the A-centered cuts (88 eV) and the top two panels, the $\Gamma$-centered cuts (108 eV). (c) A- and $\Gamma$-centered Fermi Surface maps (bottom and top panels, respectively).  The red dashed line is drawn along the direction of the EDM, which is aligned along the $\Gamma$K(AH) direction. Hexagonal 2D projection of the BZ is outlined by the dashed white hexagon. (d) Zoom in the region around $\Gamma$ and A points, with the respective second derivative plot. Dashed lines are guides to the eye for band visualization.}
\end{figure*}

As shown in Figure \ref{fig:1}(a), the material adopts a layered crystal structure (trigonal space group $P3m1$, number $164$), wherein the Te atoms form hexagonal lattices sandwiching a layer of Zr atoms. The presence of SOC, a quasi-2D hexagonal lattice and bands derived from $p$-orbitals make ZrTe$_{2}$ a type-II Dirac semimetal candidate \cite{nguyen_fermiology_2022, tian_topological_2020, tsipas_massless_2018}. This system has also been demonstrated to be a platform for studying electronic orders, such as charge density waves \cite{song_signatures_2023, zhang_emergent_2023} and superconductivity \cite{machado_evidence_2017,correa_superconductivity_2022, zhang_electronic_2020,PhysRevMaterials.9.064802}.

Experimental results by L. E. Correa \textit{et al.} (Ref. \cite{correa_evidence_2022}) suggest that Ni-intercalation drives an instability in Ni$_{x}$ZrTe$_{2}$ at $T^{*}$ $\approx 287$ K, which precedes a superconducting transition at a much lower temperature (T$_{C} = 4$ K). The instability is characterized by a clear kink in the resistivity $\rho(T)$ and by a jump in the heat capacity (C$_{p}$), both at T$^{*}$. These are typical signatures of electronic and/or structural transitions, the nature of which is not clear. 

Motivated by these experiments, here we present a high-resolution ARPES study of $1T$-ZrTe$_2$ and Ni$_{x}$ZrTe$_{2}$. Our results for the ZrTe$_2$ electronic structure are in excellent agreement with previous detailed ARPES experiments and \textit{ab initio} density functional theory (DFT) calculations of this compound  \cite{correa_evidence_2022,kar_metal-chalcogen_2020}. For the intercalated material, flat bands with binding energies E$_{B} \approx -0.7$ eV and E$_{B} \approx -1.2$ eV are clearly observed and are associated with the sharp increase in the electronic density of states (DOS) predicted from DFT calculations \cite{correa_evidence_2022}. Furthermore, and most strikingly, we characterize a reconstruction of the electronic structure at low $T$ which halves the periodicity of the electronic bands along the $k_{z}$ direction. This is reminiscent of a band folding driven by a $\boldsymbol{q}=(0,0,\pi)$ wave-vector. Supported by theoretical calculations, we ascribe this result to a charge density wave (CDW) transition driven by interband hole-hole interactions, due to a band crossing which takes place above the Fermi level (E$_{\text{F}}$).

\section{METHODS}\label{sec:methods}

ZrTe$_{2}$ and Ni$_{x}$ZrTe$_{2}$  single crystals were grown by isothermal chemical vapor transport (ICVT) \cite{correa_growth_2022}. Their structure and composition were characterized by X-ray diffraction (XRD), energy-dispersive X-ray spectroscopy (EDS) and inductively coupled plasma (ICP). Overall, the amount of Ni is relatively small with $x\approx0.05$. Electrical transport and heat capacity measurements were performed in fresh samples synthesized specifically for this ARPES investigation, using a Physical Properties Measurements System (PPMS) from Quantum Design. Results are discussed in the supplemental material Ref. \cite{suppmat} and essentially reproduce the findings of  Ref.  \cite{correa_evidence_2022}.

The ARPES experiments were conducted at the Bloch beamline of the Max IV synchrotron facility, wherein data were collected by a high-performance DA30 hemispherical electron analyzer from Scienta Omicron. The measurements were carried out along the high symmetry directions $\Gamma$M/AL and $\Gamma$K/AH for two photon energies, corresponding to both the $108$ eV ($\Gamma$) and $88$ eV (A) points in the Brillouin Zone (BZ). An extra set of measurements at $40$ eV at $20$ K was also carried out for the Ni-doped sample. Both $\pi$ (linear horizontal) and $\sigma$ (linear vertical) polarized X-rays were used to probe different Te-$4p$ and Zr-$4d$ orbital contributions to the electronic band structure. The total energy resolution is approximately $8$ to $10$ meV, and the experimental setup provided an angular resolution of $0.1^\circ$. Photon energies were selected based on the $k_z$-dispersion investigation of the high symmetry points positioning \cite{suppmat}. The  E$_{\text{F}}$ determination is also found in Ref. \cite{suppmat}. 

Samples were cleaved in the main preparation chamber, with a vacuum better
than $4 \times 10^{-9}$ mbar, using either aluminum posts or Scotch tape. Following cleavage, the samples were transferred to the analyzer chamber, where the vacuum was maintained below $3 \times 10^{-11}$ mbar for measurements. Both samples were measured at $20$ K, the lowest temperature achievable at the beamline and well below $T^*$. For the Ni-intercalated sample, additional measurements were carried out at $290$ K and $100$ K, corresponding to temperatures above and below $T^*$, respectively. Temperature control was achieved using a six-axis cryostat manipulator equipped with a closed-cycle liquid helium system.

\begin{figure*}
\includegraphics[width=\linewidth]{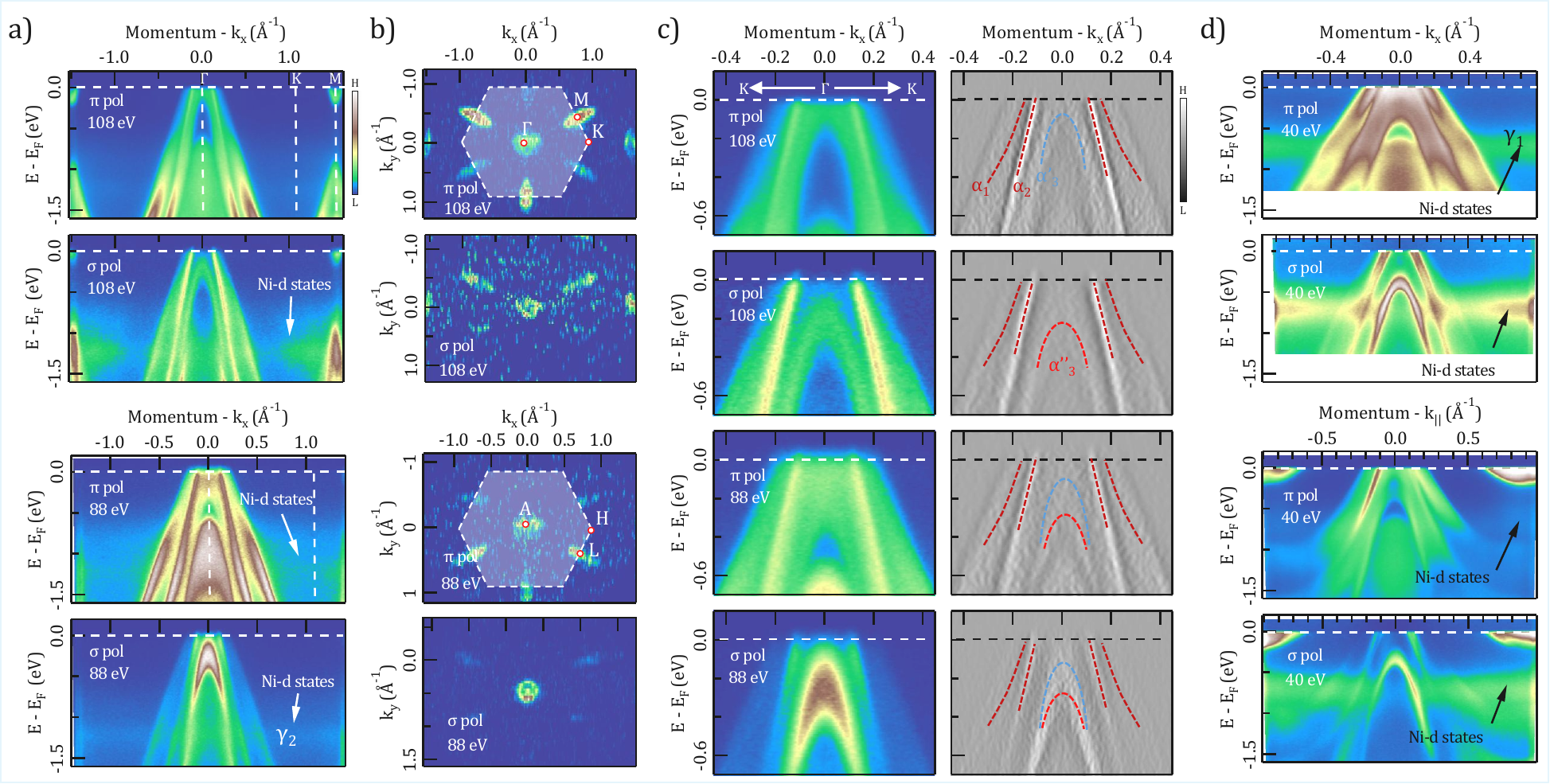}
\caption{\label{fig:2} (a) EDMs for $\Gamma$K high symmetry direction at both photon polarization and BZ planes for the Ni-intercalated sample at $T=20$ K. The bottom two maps display the AH cut ($88$ eV), and the top two maps display the $\Gamma$K cut ($108$ eV). (b) Fermi Surface maps for the respective EDMs. The dashed white hexagon outlines the hexagonal 2D projection of the BZ. (c) Zoom in the region around $\Gamma$/A, with the respective second derivative plot. Dashed lines are eye-guides to the band visualization. Notably, the $\alpha^{'}_{3}$ band is absent at the A plane. (d) EDMs taken with $40$ eV photon energy. From these maps, it is possible to observe the emergence of another flat band state at E$_B \approx -0.7$ eV, which is not evident in the higher energy maps. The top two maps are taken along the $\Gamma$K direction and the bottom two along the $\Gamma$M.}
\end{figure*}

\section{RESULTS AND DISCUSSION}

\subsection{\label{sec:ZrTe2}Parent Compound}

Figure \ref{fig:1} shows ARPES data obtained for ZrTe$_2$. An overview of the electronic band structure is provided in the panels of Figure \ref{fig:1}(b), which display the Energy Distribution Maps (EDMs) taken along $\Gamma$-K-M and A-H-L high symmetry directions of the hexagonal BZ. Upper panels are dedicated to EDMs centered about $\Gamma$ and the lower panels for EDMs centered about A. All main features of the electronic band structure can be observed: for $\Gamma$-centered measurements, direct inspection of the data suggests that two hole-like bands cross E$_{\text{F}}$ at $\Gamma$ and one electron-like band crosses E$_{\text{F}}$ at M. As for A-centered measurements, the situation is similar, but the crossing of one of the hole-like bands at A cannot be resolved. 

We then investigate the EDMs for a limited range in binding energy, as shown in the panels of Figure \ref{fig:1}(d). With the assistance of the respective second derivatives, we observe four hole-like bands, denoted henceforth as $\alpha_1$, $\alpha_2$, $\alpha^{'}_3$ and $\alpha^{''}_3$. The $\alpha_1$ band crosses E$_{\text{F}}$ and is visible for both light polarizations. On the other hand, the $\alpha^{'}_3$ and $\alpha^{''}_3$ bands do not cross the Fermi level and their intensities are polarization dependent: the $\alpha^{'}_3$ band intensity is stronger for $\pi$-polarized light, whereas  $\alpha^{''}_3$ is stronger for $\sigma$-polarized light. At $\Gamma$, the $\alpha_2$ band clearly crosses E$_{\text{F}}$, but at A it barely contributes to the density of states at E$_{\text{F}}$. This is the main difference between the $\Gamma$- and A-centered electronic structures.  Far from the BZ center, one electron-like band, $\beta$, forming an electron pocket, is observed about the M(L) high symmetry points. This  $\beta$ band is slightly stronger for $\pi$-polarized light.

Overall, the polarization dependence of the bands forming the hole pockets hints at their orbital character. Unambiguous orbital character assignment of the electronic bands in ZrTe$_{2}$, however, is far from trivial because of strong SOC. For instance, the $\alpha_2$ and $\beta$ bands, in DFT calculations without SOC \cite{correa_evidence_2022}, derive from Te $p_{z}$ and Zr $d_{ z^2}$ orbitals, respectively. In the experiments, these bands are stronger for $\pi$-polarized light but are clearly observed for $\sigma$-polarized light, showcasing the mixing of orbital character due to SOC.

Our results are in good agreement with the mentioned DFT calculations and previous ARPES studies of bulk ZrTe$_{2}$ \cite{kar_metal-chalcogen_2020, zhang_electronic_2020}. It is important to note, however, that the $\alpha_{3}$ band pair is not captured in DFT calculations, which may indicate that these are surface states.  Also, we demonstrate that bands $\alpha_{2}$ and $\alpha^{'}_{3}$ are indeed two different states, which was not clear from previous studies. 

Regarding the Fermi Surface (FS) Maps, the k$_x$-k$_y$ plane is shown in panel Figure \ref{fig:1}(c). These maps show the hexagonal symmetry inherent to the system, a characteristic further reinforced by the overlay of the two-dimensional BZ drawn on top of the maps. Overall, the FS consists of one electron pocket located at the M(L) point and two hole pockets around the $\Gamma$ point. The inspection of the FS maps alone does not allow distinguishing the presence of either one or two hole pockets around the A point. Our second derivative analysis suggests only one hole pocket around A, in contrast with the clearly two distinguished pockets around $\Gamma$. Interestingly, the electron pocket exhibits a distinctive three-fold intensity symmetry, which makes the K and K’ nonequivalent points.

\begin{figure*}[t!]
\includegraphics[width=\linewidth]{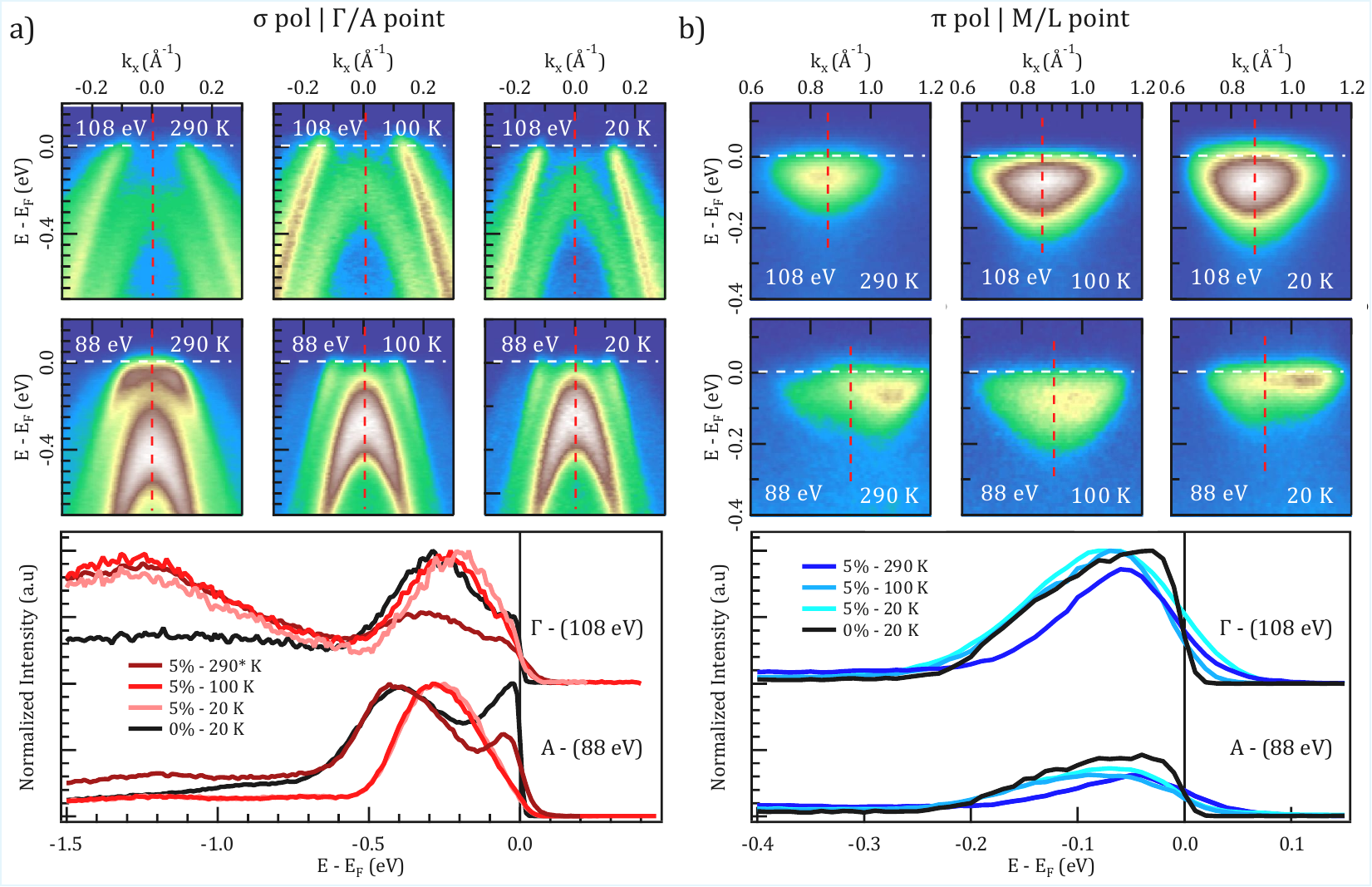}
\caption{\label{fig:3} (a) Zoom into the band structure around the A ($88$ eV) and $\Gamma$ ($108$ eV) points for the Ni-intercalated sample at temperatures of $290$, $100$ and $20$ K with $\sigma$-pol photons. The lower panel shows the respective EDCs (taken at the dashed red lines): around the A cut, the presence of the $\alpha_{2}$ band peak for the parent compound (black line) and T$>\text{T}^{*}$ (dark red) is notable. This peak, however, is absent for the low-temperature measurements. (b): Same as (a), but for the electron pocket around L ($88$ eV) and M ($108$ eV) points. From the EDCs we see a shift away from the Fermi Level.}
\end{figure*}

\subsection{\label{sec:NiZrTe2}Ni-intercalated Compound}

Turning to the Ni-intercalated sample, Figure \ref{fig:2}(a) shows the EDMs, obtained for $T=20$ K, for $88$ (bottom panels) and $108$ (upper panels) eV along the $\Gamma$-K-M and A-H-L directions. The most salient feature introduced by Ni intercalation is the emergence of two flat-like bands with binding energies E$_B \approx -1.2$ eV and E$_B \approx -0.7$ eV, denoted $\gamma_1$ and $\gamma_2$, respectively. DFT calculations \cite{correa_evidence_2022} proposed that Ni intercalation introduces sharp and large contributions, derived from Ni $d$-orbitals, to the DOS in this same energy region, suggesting that this is the origin of the $\gamma_1$ and $\gamma_2$ weakly dispersive features.

Although the flat band $\gamma_2$ displays very clear intensity for high-energy measurements ($88$/$108$ eV), $\gamma_1$ cannot be well distinguished from those energy maps. Upon further inspection with lower incident photon energy ($40$ eV), we could observe the clear spectral weight from this emerging flat band, as shown in Figure \ref{fig:2}(d). In Figure \ref{fig:2}(b), the $\Gamma$ and A-centered FS Maps are presented. 


Closer inspection of the electronic bands in the vicinity of E$_{\text{F}}$ then leads to our main finding. For a limited binding energy interval, Figure \ref{fig:2}(c) shows the EDMs around A (bottom panels) and $\Gamma$ (upper panels) with both light polarizations. For the maps at $\Gamma$ (108 eV), we can see clearly the presence of both $\alpha'_{3}$ and $\alpha''_{3}$, upon changing the photon polarization. Surprisingly, after measuring at A (88 eV), the same features are found, as opposed to the parent compound case. In reference to the second derivatives (right panels of Figure \ref{fig:2}(c)), it becomes clear that a second  band also crosses E$_{\text{F}}$ at A for the Ni-intercalated sample. 

\begin{figure*}[t!]
\includegraphics[width=0.95\linewidth]{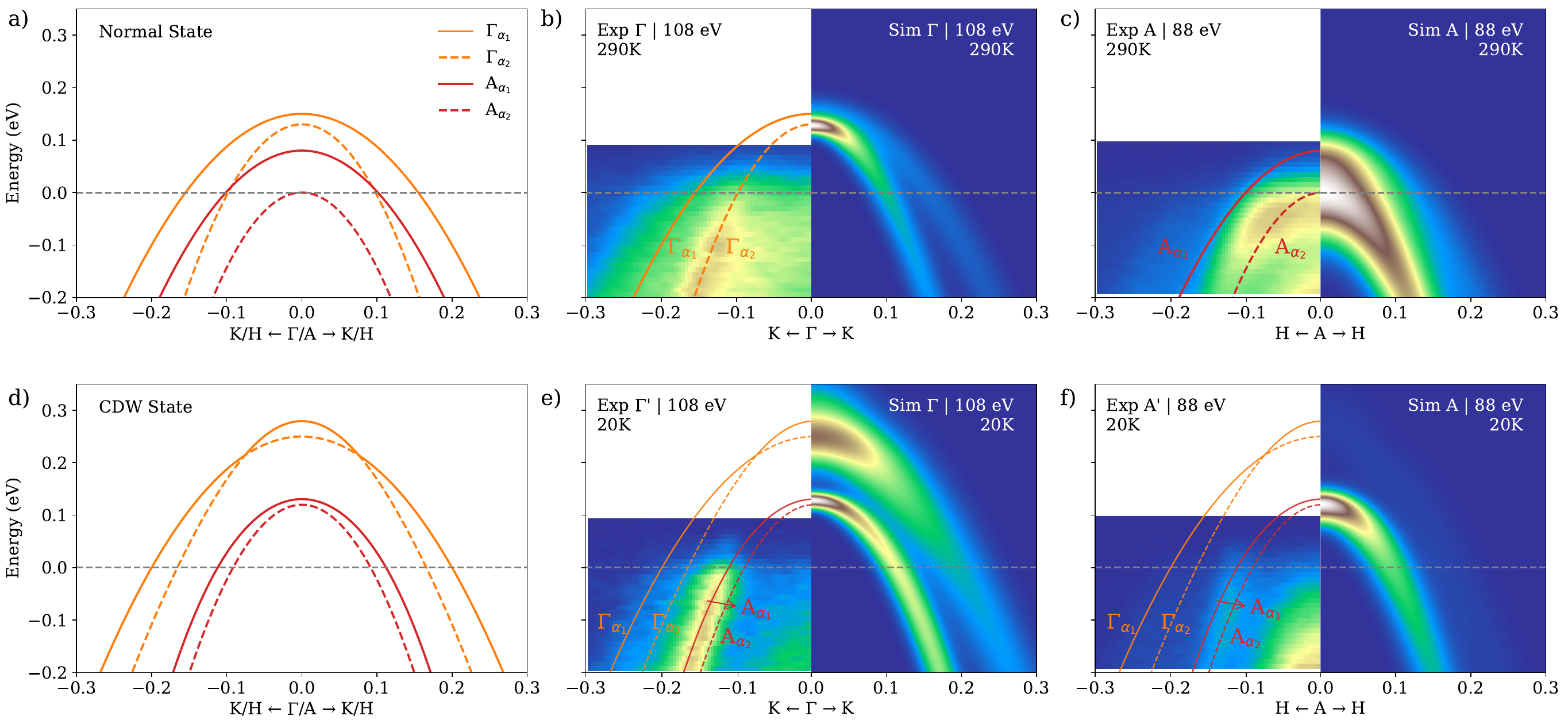}
\caption{Illustration of the electronic band reconstruction scenario for a CDW state. (a) Normal state electronic band structure showing the two $\Gamma$- and A-centered $\alpha$ bands. (b) Split panel showing the experimental data (left) and simulated spectral weight (right) for the normal state $\Gamma$-centered electronic band structure. (c) Same as (b), but for the A symmetry point. (d) CDW reconstructed electronic band structure, showing the resulting folded bands. (e) Split panel showing the experimental data (left) and simulated spectral weight (right) for the CDW state band structure in $\Gamma$. (f) Same as (e), but for the A symmetry point.}\label{fig:model}
\end{figure*}

To further investigate this finding, we performed $T$-dependent ARPES experiments for the Ni-intercalated sample at the $\Gamma$ and A points. Figure \ref{fig:3}(a) shows the EDMs around the A and $\Gamma$ points with $\sigma$-pol for $T=290$, $100$ and $20$ K, alongside the Energy Distribution Curves (EDCs) cuts at the BZ center. First, we note that the electronic band structure around $\Gamma$ and A, at $T = 290 > T^{*} = 287$ K for Ni$_x$ZrTe$_2$, is very similar to the one for ZrTe$_2$ at $T = 20$ K, apart from intensities. This is supported by the EDCs in the bottom panels of Figure \ref{fig:3}(a). Indeed, for $T = 290$ K, there is a key difference between the $\Gamma$- and A-centered electronic structures for both samples: at $\Gamma$, the spectral weight close to E$_F$ comes mainly from the $\alpha_{2}$ band, which crosses E$_{\text{F}}$ with a finite $k_F$. At A, the peaks closer to E$_{\text{F}}$ are the $\alpha_2$ band maxima, which do not cross E$_{\text{F}}$. Upon lowering $T$, at $T = 100$ K, one can observe from EDMs and EDCs that the maximum of the $\alpha_{2}$ band at the A plane is removed from the vicinity of E$_F$ in the Ni$_x$ZrTe$_2$ case. The $\alpha_{2}$  band maximum is now above E$_F$, as in the case of the $\Gamma$-centered electronic structure. The same picture holds for measurements at $T = 20$ K. These observations further characterize the Ni$_x$ZrTe$_2$ electronic band structure reconstruction at low temperatures, reflecting that the periodicity along the $k_{z}$ direction was halved.

In the monolayer ZrTe$_2$ system, a band folding of $\Gamma$ into M was observed for T $< 150$ K \cite{song_signatures_2023}. We then inspect the electronic structure around the M point in Figure \ref{fig:3}(b), where we present EDMs for the electron pockets around M/L and the pocket-centered EDCs, for the three $T$s. Apart from the differences in spectral weight, lowering $T$ leads to an apparent band shift away from the Fermi Level, stronger at the lowest temperature. Whereas this shift is characteristic of a gap opening typical of CDW phase transitions, no other signs of electronic reconstruction could be resolved by our measurements. We therefore conclude that no in-plane electronic reconstruction is taking place in our system.

The outstanding task is thus to present a mechanism for the electronic structure reconstruction along the $k_{z}$ direction. This motivates us to look at a simplified model for a putative CDW state driven by electronic interactions, similarly considered in the case of $1T$-TiSe$_2$ \cite{monney_spontaneous_2009, monney_exciton_2011,monney_probing_2010} (as well as Refs. \cite{suppmat,aebi_search_2001,zhu_misconceptions_2017,zhu_classification_2015}). Within a mean-field theory, the CDW is characterized by a nesting vector $\boldsymbol{Q}=(0,0,\pi)$, which folds the electronic structure overlapping bands from different regions of the BZ. In particular, the nesting occurs by interband hole-hole interactions mixing the normal state $\Gamma_{\alpha_{2}}$ and $A_{\alpha_{1}}$ bands. The CDW gap opens locally at the band crossings, as shown in Figure \ref{fig:model}(a), leading to avoided crossings rather than a global gap. In this regime, the folded bands acquire significant spectral weight near the crossing energies, while their intensity rapidly diminishes for energies away from these points. 

The model was solved assuming a parabolic parametrization of the experimentally observed bands about the A and $\Gamma$ points, which should be valid within a narrow region close to E$_F$. The parameters were extracted from MDCs of the high-temperature spectra (representative of the normal-state electronic structure). For the CDW strength, we adopt a CDW order parameter ($\Delta$) of $70$ meV. In Figure \ref{fig:model}(a), the normal-state electronic band structure near E$_F$ at $\Gamma$/A is shown, alongside the corresponding experimental and simulated data in Figs. \ref{fig:model}(b) and (c). In Fig. \ref{fig:model}(d), the schematic electronic reconstruction of $\Gamma$ into A is shown, without taking into consideration the spectral weight renormalization. In Figs. \ref{fig:model}(e)-(f), the panels present, on the left, the experimental data and, on the right,  the respective simulated spectra for the $\Gamma$- and A-centered electronic structures. This time, the spectral weight renormalization and experimental broadening are included. In Figs. \ref{fig:model}(d)-(f), the band labels $\Gamma_{\alpha_{2}}$ and $A_{\alpha_{1}}$ intend to provide a guide upon which one can deduce the main original band contributing to that particular mixed band within the CDW state. The bands in the ordered state are shifted by about $100$ meV, which is $\approx\Delta$, as adopted in the $1T$-TiSe$_2$ case \cite{monney_spontaneous_2009, monney_exciton_2011,monney_probing_2010}. This shift is intended to reproduce the change of the chemical potential in the ordered state. 

Comparison between theory and experiments suggests that our picture, meaning a CDW driven by interband hole-hole interactions, provides a minimal model describing the observed phenomenology.  In detail, as Fig. \ref{fig:model}(a) illustrates, the band crossing occurs above the E$_F$, which averts the direct detection of the gap by ARPES and causes the redistribution of spectral weights in the region below  E$_F$.  The $\Gamma$-centered electronic structure features one strong band forming the inner hole pocket, which derives mainly from mixed $\Gamma_{\alpha_{2}}$ and $A_{\alpha_{1}}$ bands (see arrow in  Fig.\ref{fig:model}(e)). The weak band forming the outer-hole pocket is the shifted $\Gamma_{\alpha_{1}}$ band of the normal state. Other bands have zero spectral weight in this region. In the case of the A-centered electronic structure, the main spectral weight is due to the shifted  $A_{\alpha_{2}}$ band (see arrow in Fig.\ref{fig:model}(f)). The spectral weight of the mixed $\Gamma_{\alpha_{2}}$ and  $A_{\alpha_{1}}$ goes to zero when approaching E$_F$. As the simulation proposes, the band forming the outer hole pocket is very weak, but it can be observed for larger binding energies as shown in Fig. \ref{fig:2}.


Overall, our model explains the equivalence between the $\Gamma$- and A-centered electronic bands at low-$T$. In this sense, we unveil a mechanism where the CDW gap opens above E$_F$ and is driven by the nesting between the hole-like $\Gamma_{\alpha_{2}}$ and  $A_{\alpha_{1}}$ electronic bands. 

The onset of a CDW phase leads to a structural transition \cite{zhu_misconceptions_2017,zhu_classification_2015}. Preliminary  X-ray diffraction data of a Ni$_x$ZrTe$_2$ single crystal, probing $(00l)$ reflections, measured at $T = 300$ K and $T=80$ K, display no extra reflections due to this possible lattice reconstruction \cite{suppmat}. Further structural investigation of Ni$_x$ZrTe$_2$ is certainly motivated by these results.


We finally mention that both at the beamline and in He-lamp ARPES results, and for different samples, we observed a gradual shift in the chemical potential $\mu$, as the system was cooled down. This could be related to the fact that $\mu$ becomes strongly $T$-dependent in the CDW phase, in particular for a transition of relatively high $T^{*}= 287$ K. This shift persists even after waiting a few minutes for the system to reach thermal stability, in the form of a time-dependent shift. This is also reminiscent of a charging effect, but the observed change in electronic mobility is not fully compatible with this picture. In our measurements, one may notice that the EDC data exhibit an unusual behavior: at low $T$, the spectra do not display a sharp Fermi–Dirac cutoff, as would normally be expected. Indeed, since each spectrum required approximately 10 minutes to acquire, the time-dependent shift in $\mu$ manifests in the data as an overall broadening of the Fermi edge ($\approx 25$ meV, as explained in the Supp. Material \cite{suppmat}). This effect was not observed in measurements of other intercalated or defect-engineered ZrTe$_2$ materials, such as Dy-intercalated  ZrTe$_2$ \citep{moya_estrutura_2023}.

\section{Summary and Conclusions}

In our work, we investigated the electronic band structures of ZrTe$_2$ and Ni$_{x}$ZrTe$_2$. In the case of ZrTe$_2$, results are in agreement with previous experimental investigations and DFT calculations. Our low-$T$ investigation confirms that the electronic band structure is nearly $T$-independent. In the case of Ni$_{x}$ZrTe$_2$, we observed two flat band states, likely derived from the localized Ni 3$d$ state, as supported by previous DFT calculations. 

Most importantly, for $T < 100$ K ($< T^{*}= 287$ K), an electronic reconstruction that halves the periodicity of the electronic bands along $k_{z}$ was identified in Ni$_{x}$ZrTe$_2$. We then proposed a model where interband hole-hole interactions trigger the electronic structure reconstruction leading to a putative CDW state. This model was solved taking into account the parametrization of the experimentally determined electronic bands and assuming a pairing interaction of $\Delta = 70$ meV, compatible with the observed $T^{*}= 287$ K energy scale. Comparison between theoretical modeling and experimental data suggests that our model captures the observed phenomenology.

\begin{acknowledgments}
We acknowledge MAX IV Laboratory for beamtime on the Bloch Beamline under Proposal 20210928. The support of Jacek Osiecki and Craig Polley throughout the beamtime was fundamental for the success of the measurements. Research conducted at MAX IV, a Swedish national user facility, is supported by the Swedish Research council under contract 2018-07152, the Swedish Governmental Agency for Innovation Systems under contract 2018-04969, and Formas under contract 2019-02496. We acknowledge financial support from CNPq Grant Nos. 304165/2023-9 (A.J.S.M.) and 307686/2026-4 (F.A.G.); FAPESP, Grants Nos. 2019/05150-7 (M.R.C), 2023/14902-8  (L.B.B.), 2024/13291-8 (F.A.G.), 2021/14322-6 and 2018/08819-2 (A.J.S.M.). This study was financed (P.H.A.M) in part by CAPES – Finance Code 001. L.B.B. acknowledges the hospitality of the Theoretical Solid State Physics Group at Ruhr-Universit\"at Bochum. 
We acknowledge the support of the INCT project Advanced Quantum Materials, involving the Brazilian agencies CNPq (Grant No, 408766/2024-7), FAPESP (Grant No. 2025/27091-3), and CAPES.

\end{acknowledgments}

\bibliography{Ni_ZrTe2_ARPES_references}

\end{document}